\documentclass[hyper]{JHEP3}
\usepackage[dvips]{graphicx}
\usepackage{latexsym,amsmath,amssymb}

\title{Thermodynamic similarity between the noncommutative
  Schwarzschild black hole and the Reissner-Nordstr\"om black hole} 

\author{Wontae Kim \\
  Department of Physics and Center for Quantum Spacetime, Sogang University,
  Seoul 121-742, Korea \\
  E-mail: \email{wtkim@sogang.ac.kr}}
\author{Edwin J. Son \\
  Department of Physics and Basic Science Research Institute, Sogang
  University, Seoul 121-742, Korea \\ 
  E-mail: \email{eddy@sogang.ac.kr}}
\author{Myungseok Yoon \\ 
  Center for Quantum Spacetime, Sogang University, Seoul 121-742,
  Korea \\
  E-mail: \email{younms@sogang.ac.kr}}

\date{\today}

\abstract{%
We study thermodynamic quantities and examine the stability of a
black hole in a cavity inspired by the noncommutative geometry. It
turns out that thermodynamic behavior of the noncommutative black hole is
analogous to that of the
Reissner-Nordstr\"om black hole in the near extremal limit.
Moreover, we identify the noncommutative parameter with the squared
electric charge with some constants.}


\begin{document}

\section{Introduction}
\label{sec:intro}

Thermodynamics of a black hole is one of the most interesting issues
in the theoretical physics. Bekenstein has suggested that the entropy of a
black hole is proportional to its surface area \cite{bekenstein} and
Hawking's analysis for its origin from the point of view of quantum
field theory \cite{hawking} has led to the result that the black hole
has a thermal radiation with the temperature $T_H = \kappa/2\pi$, where
$\kappa$ is its surface gravity. Since then, there are many
studies for thermodynamics in a cavity with a finite size in various
black holes \cite{gh,hi,bcm,allen,york,wy,brown}. 

A complete explanation for the final state after the evaporation of the
black hole is important but it has not been achieved yet since the
full quantum gravity has been still unknown. However, there are two
candidates for quantum gravity, which are the string theory and the
loop quantum gravity. By the string/black hole correspondence
principle \cite{susskind}, stringy effects cannot be neglected in the
late stage of a black hole. In the string theory, coordinates of the
target spacetime become \textit{noncommutating} operators on a
\textit{D}-brane as $[\mathbf{x}^\mu, \mathbf{x}^\nu] = i
\theta^{\mu\nu} $ \cite{witten}, where $\theta^{\mu\nu}$ is an
anti-symmetric matrix which determines the fundamental cell
discretization of spacetime much in the same way as the Planck
constant $\hbar$ discretizes the phase space. Recently, it has been
shown that Lorentz invariance and unitary, raised in the
Weyl-Wigner-Moyal *-product approach, can be achieved by assuming
$\theta^{\mu\nu} = \theta\ \mathrm{diag} (\epsilon_1, \cdots,
\epsilon_{D/2})$ \cite{ss,nss,anss}, where $\theta$ and $D$ are a
constant and the dimension of spacetime \cite{ss1}. In Ref.~\cite{nf}.
there has been the study on the thermodynamics of the
Reissner-Nordstr\"om (RN) black hole, considering the effects of space
noncommutativity.

In this work, we would like to study thermodynamics of a static and
spherically symmetric black hole, considering the effects of
noncommutative geometry. Especially, we wish to point out an analogy
between the noncommutative black hole and the RN black hole, which has
been commented shortly in Ref.~\cite{mkp:rbh}. We shall show that the
parameter $\theta$ in the noncommutative black hole plays a similar
role with an electric charge in RN black hole. In Sec.~\ref{sec:NBH},
we introduce a noncommutative black hole and examine the relation
between the mass and its horizons. Comparing the noncommutative black
hole to the RN black hole, it can be shown that there is an analogy
between them in Sec.~\ref{sec:extremal}. In Sec.~\ref{sec:thermo}, we
analyze the thermodynamic properties of the noncommutative black hole
in a cavity with a finite size and check the thermodynamic stability
of the black hole. It can be found that these properties are similar
to those of the RN black hole. Finally, some discussions are given in
Sec.~\ref{sec:dis}.

\section{Schwarzschild black hole inspired by the noncommutative geometry}
\label{sec:NBH}

We would like to examine the metric of the Schwarzschild black hole
when there exists the noncommutativity of spacetime.
It has been shown that noncommutativity eliminates point-like
structures in favor of smeared objects in flat
spacetime~\cite{ss}. The effect of smearing is mathematically
implemented by replacing the position Dirac-delta function with a
Gaussian distribution of the width $\sqrt{\theta}$. In a static,
spherically symmetric case with this logical connection, the mass
density of a gravitational source is chosen as~\cite{nss}
\begin{equation}
  \label{rho}
  \rho_\theta = \frac{M}{(4\pi\theta)^{3/2}} \exp \left(
    -\frac{r^2}{4\theta} \right),
\end{equation}
where the total mass $M$ is diffused throughout the region of linear
size $\sqrt{\theta}$ and the $\theta$ is a constant parameter representing
noncommutativity.

For a static and spherically symmetric metric, the density~(\ref{rho})
and the conservation law tell us that the energy-momentum tensor is
given by ${T^\mu}_\nu = {\rm diag} (-\rho_\theta, p_r, p_\bot, p_\bot
)$, where the radial and the tangential pressure are given by $p_r =
-\rho_\theta$ and $p_\bot = -\rho_\theta - \frac12 r \partial_r
\rho_\theta$, respectively. Then, solving the Einstein equations of
motion, we obtain the line element as 
\begin{equation}
  \label{line}
  ds^2 = -f(r) dt^2 + f(r)^{-1} dr^2 + r^2 d\Omega_2^2
\end{equation}
with 
\begin{equation}
  \label{f}
  f(r) = 1-\frac{2m(r)}{r} = 1 - \frac{4M}{r\sqrt{\pi}} \gamma\left( \frac32,
    \frac{r^2}{4\theta} \right),
\end{equation}
where the mass distribution
$m(r)=(2M/\sqrt{\pi})\gamma(3/2,r^2/4\theta)$ is straightforwardly
obtained from the density~(\ref{rho}), and the lower incomplete gamma
function is defined by 
\begin{equation}
  \label{gamma}
  \gamma\left(a, z \right) \equiv
  \int_0^z t^{a-1} e^{-t} dt.
\end{equation}
Note that we only change the point-like structure of the Schwarzschild
black hole to a smeared object so that the red-shift
function~(\ref{f}) has a similar form to Schwarzschild metric except
the mass distribution $m(r)$. Moreover, it can be easily checked that
Eq.~(\ref{f}) is reduced to the Schwarzschild metric in the limit of
$r/\sqrt{\theta} \to \infty$, that is, $m(r)\to M$. 

\begin{figure}[pt]
  \includegraphics[width=0.5\textwidth]{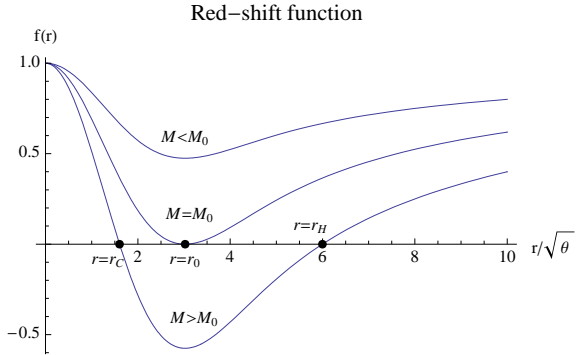}
  \caption{The red-shift function $f(r)$ is shown with respect to
    $r/\sqrt{\theta}$. There is no horizon for
    $M=\sqrt{\theta}<M_0$(top), while one degenerate and two horizons
    exist for $M=M_0\approx1.9\sqrt{\theta}$(middle) and
    $M=3\sqrt{\theta}>M_0$(bottom), respectively.} 
  \label{fig:metric}
\end{figure}

\begin{figure}[pt]
  \includegraphics[width=0.5\textwidth]{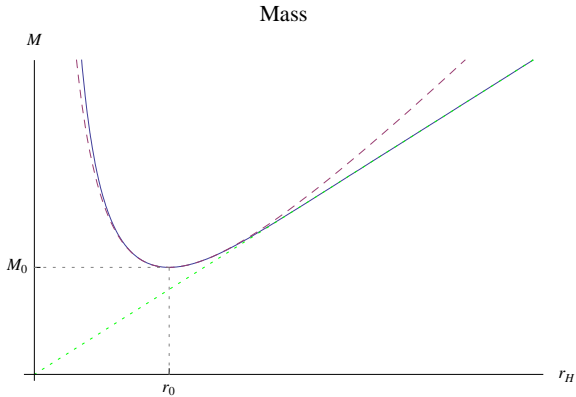}
  \caption{The solid, the dashed, and the dotted lines show the 
        relations between the mass and the horizon of the
        noncommutative, the scaled RN, and the Schwarzschild black
        holes, respectively, where ``scaled'' means $r_H = (\alpha^2-1)
        \tilde r_+$ and $\tilde r$  indicates the radial coordinate in
        the RN black hole.
      }
  \label{fig:M}
\end{figure}

However, the presence of noncommutativity changes the
Schwarzschild-like behavior into the RN-like one in the region where
the noncommutativity cannot be neglected, which will be dealt with in
detail in the next section. As a short preview, we see that there are
two horizons, \textit{i.e.}, the inner (Cauchy) horizon $r_C$ and the
outer (event) horizon $r_H$, and there exists the minimal mass $M_0$
below which no black hole can be formed. Those facts are 
seen from Fig.~\ref{fig:metric}. Moreover, at the minimal mass
$M=M_0$, the inner and the outer horizons are met at the minimal
horizon $r_0$ ($r_C \le r_0 \le r_H$).
The minimal mass is more explicitly seen in the mass relation as follows:
It results from $r_H =2m(r_H)$ that the total mass is related to the
event horizon by
\begin{equation}
  \label{M}
  M = \frac{r_H \sqrt{\pi}}{4\gamma_H},
\end{equation}
where $\gamma_H = \gamma\left(\frac32, \frac{r_H^2}{4\theta} \right)$.
Then, the minimal mass $M_0$ is explicitly seen in Fig.~\ref{fig:M}.

It seems to be appropriate to note that we can hide the parameter $\theta$
in the red-shift function by redefining mass and the radial coordinate as
$M \to M'=M/(2\sqrt\theta)$ and $r \to r'=r/(2\sqrt\theta)$ so that
the function~(\ref{f}) is reduced to $f(r) = 1 -
4M'\gamma(3/2,r'^2)/(r'\sqrt\pi)$. Then, it is deduced from the mass
relation $M'=r'_H\sqrt\pi/[4\gamma(3/2,{r'}_H^2)]$ that both the
radius and the mass of the minimal black hole are proportional to
$\sqrt\theta$ and can be written as $r_0 = 2 \alpha \sqrt{\theta}$ and
$M_0 = \sqrt{\theta\pi}/(4\alpha^2e^{-\alpha^2})$, where the constant
$\alpha \equiv r'_0$ is determined by
\begin{equation}
  \label{alpha}
  2\alpha^3 e^{-\alpha^2} = \gamma\left(\frac32, \alpha^2\right),
\end{equation}
and we can find $\alpha \approx 1.51122$ numerically, so we get
$r_0\approx3.02244\sqrt\theta$ and $M_0\approx1.90412\sqrt\theta$.

\section{Near extremal limit and analogy with the RN black hole}
\label{sec:extremal}
In spite of the similarity of the metric~(\ref{f}) to the
Schwarzschild black hole, one might think from Fig.~\ref{fig:M} that
the mass profile of the noncommutative black hole has a similar behavior
to that of the RN black hole in the vicinity of the minimal horizon
$r_0$. Since the extremal limit for the RN black hole is given by 
$\tilde M \to \tilde Q$ or $\tilde{r}_+ \to \tilde{Q}$, the extremal
limit for the noncommutative black hole can be taken as $r_H \to r_0$,
in which case $r_C$ also goes to $r_0$. 
So, this section is mainly devoted to the analogy with the RN black hole in
the near extremal limit. At this purpose, we first recall the Hawking
temperature of the RN black hole. The metric of the RN black hole is given by
\begin{equation}
  \label{metric:RN}
  \tilde{f}(\tilde{r}) = 1- \frac{2\tilde{M}}{\tilde{r}} +
  \frac{\tilde{Q}^2}{\tilde{r}^2}, 
\end{equation}
where $\tilde{M}$ and $\tilde{Q}$ are the mass and the electric charge of the
black hole. The inner ($r_-$) and outer ($r_+$) horizons are given by
$\tilde{r}_\pm = \tilde{M} \pm \sqrt{\tilde{M}^2 - \tilde{Q}^2}$. 
Since the mass and the charge can be written as $\tilde{M} =
(\tilde{r}_+ + \tilde{r}_-)/2$ and $\tilde{Q} = \sqrt{\tilde{r}_+
  \tilde{r}_-}$, we can rewrite the mass in terms of $\tilde{r}_+$ and
$\tilde{Q}$, similarly to Eq.~(\ref{M}), 
\begin{equation}
  \label{M:RN}
  \tilde{M} = \frac12 \left(\tilde{r}_+ +
    \frac{\tilde{Q}^2}{\tilde{r}_+} \right) \ge \tilde{M}_0 =
  \tilde{Q}, 
\end{equation}
where the equality is satisfied with $\tilde{r}_+ = \tilde{r}_0 =
\tilde{Q}$. The Hawking temperature is obtained as
\begin{equation}
  \label{TH:RN}
  T^{\rm RN}_H = \frac{\tilde{r}_+ - \tilde{r}_-}{4\pi \tilde{r}_+^2}
  = \frac{\tilde{r}_+^2 - \tilde{Q}^2}{4\pi \tilde{r}_+^3}.
\end{equation}
Then, the Hawking temperature of the near extremal RN black hole is written as 
\begin{equation}
  \label{TH:RN:ex}
  T_H^{\rm RN} \simeq \frac{\tilde{r}_+ - \tilde{Q}}{2\pi \tilde{Q}^2}
  + O(\tilde{r}_+ - \tilde{Q})^2. 
\end{equation}

On the other hand, the Hawking temperature of the noncommutative black
hole is calculated as 
\begin{equation}
  \label{TH}
  T_H = \frac{1}{4\pi r_H} \left[1 - \frac{M
      r_H^2}{\sqrt{\pi}\theta^{3/2}} \exp\left(
      -\frac{r_H^2}{4\theta}\right) \right].
\end{equation}
For the limit of $r_H \gg 2\sqrt{\theta}$, it recovers the Hawking
temperature of the Schwarzschild black hole $T_H = 1/(4\pi r_H)$.
Now, the Hawking temperature near extremal regime is given by 
\begin{equation}
  \label{TH:ex}
  T_H \simeq \frac{\xi (r_H - r_0)}{2\pi r_0^2} + O(r_H - r_0)^2,
\end{equation}
where the mass relation~(\ref{M}) is used and $\xi = \alpha^2 - 1$.

Comparing Eqs.~(\ref{TH:RN:ex}) and (\ref{TH:ex}), one can easily
find the relations between the horizons by identifying 
\begin{eqnarray}
  r_H &=& \xi \tilde{r}_+, \label{rH:r+} \\
  r_0 &\equiv& 2\alpha\sqrt{\theta} = \xi \tilde{Q}, \label{r0:Q} 
\end{eqnarray}
then, Eq.~(\ref{TH:ex}) becomes
\begin{equation}
  \label{TH:ex:sub}
  T_H \simeq \frac{\tilde{r}_+ - \tilde{Q}}{2\pi \tilde{Q}^2} +
  O(\tilde{r}_+ - \tilde{Q})^2, 
\end{equation}
which concludes $T_H \simeq T_H^{\rm RN}$ in the leading order.
Thus, the noncommutative black hole behaves similar to the RN black
hole in the near extremal limit. Moreover, the noncommutative
parameter is related to the charge of the RN black hole by
\begin{equation}
  \label{theta:Q}
  \theta = \frac{\xi^2\tilde{Q}^2}{4\alpha^2}.
\end{equation}

However, the rescaled radius
$r_H=\xi\tilde{r}_+$ does not match the two minimal masses $M_0$ and
$\tilde{M}_0$, since $r_0 \ne M_0$ whereas $\tilde{r}_0 =
\tilde{M}_0$. In order to find out the relation between the two masses
$M$ and $\tilde{M}$, we first expand Eq.~(\ref{M}) in the near
extremal limit, 
\begin{equation}
  \label{M:ex}
  M \simeq M_0 \left[ 1 + \xi \left( \frac{r_H - r_0}{r_0} \right)^2
  \right] + O(r_H-r_0)^3. 
\end{equation}
Next, Eq.~(\ref{M:RN}) is expanded in the near extremal limit as
\begin{equation}
  \label{M:RN:ex}
  \tilde{M} \simeq \tilde{Q} \left[ 1 + \frac12 \left(
      \frac{\tilde{r}_+ - \tilde{Q}}{\tilde{Q}} \right)^2 \right] +
  O(\tilde{r}_+ - \tilde{Q})^3. 
\end{equation}
So, the two masses are related in the near extremal limit as
$(M-M_0)/M_0\simeq2\xi(\tilde{M}-\tilde{Q})/\tilde{Q}$ in the leading
order. In fact, the curve for the RN black hole in Fig.~\ref{fig:M} is plotted
with respect to the rescaled radius $r_H = \xi \tilde{r}_+$.

\section{Thermodynamics of the noncommutative black hole}
\label{sec:thermo}

\begin{figure}[pt]
  \includegraphics[width=0.5\textwidth]{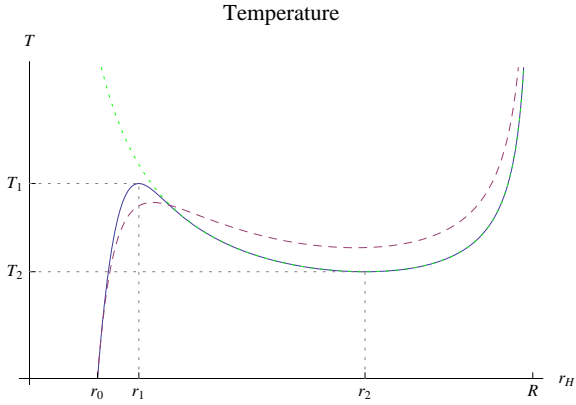}
  \caption{The solid, dashed, and dotted lines show the 
        relations between the local temperature and the horizon of the
        noncommutative, the scaled RN, and the Schwarzschild black
        holes, respectively. For $R=10$ and $\theta=0.2$, we obtain
        $r_0 \approx 1.35167$, $r_1 \approx 2.17241$, $r_2 \approx
        6.66667$, $T_1 \approx 0.0377456$, and $T_2 \approx 0.0206748$.
      }
  \label{fig:T}
\end{figure}

We now consider a cavity with the finite size $R$. The local
temperature on the boundary of the cavity is given by~\cite{bcm}
\begin{equation}
  \label{T}
  T = \frac{T_H}{\sqrt{f(R)}}.
\end{equation}
It can be seen from Fig.~\ref{fig:T} that the local temperature of
the noncommutative black hole behaves like the RN (Schwarzschild) black hole
for the small (large) black hole. 
The temperature has two extrema: one is the local maximum at $r_H = r_1$
and the other is the local minimum at $r_H = r_2$. There is one small
black hole for $0<T<T_2$ and one large black hole for $T>T_1$, where
$T_i = T|_{r_H = r_i}$ with $i = 1,2$. For the case of $T_2 < T < T_1$,
there are three black hole solutions.

\begin{figure}[pt]
  \includegraphics[width=0.5\textwidth]{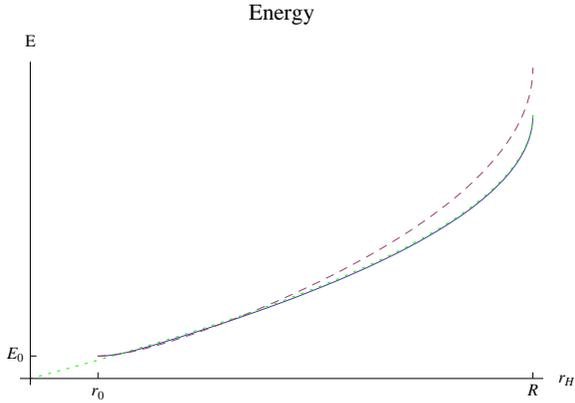}
  \caption{The solid, the dashed, and the dotted lines show the 
        relations between the energy and the horizon of the
        noncommutative, the scaled RN, and the Schwarzschild black
        holes, respectively. For $R=10$ and $\theta=0.2$, we obtain
        $r_0 \approx 1.35167$ and $E_0 \approx 0.904251$.
      }
  \label{fig:E}
\end{figure}

Since the entropy is proportional to the area of event
horizon by
\begin{equation}
  \label{S}
  S = \frac{A}{4} = \pi r_H^2,
\end{equation}
and the first law of thermodynamics $dE = TdS$ should be satisfied
for a fixed $R$, we obtain the energy as
\begin{equation}
  \label{E}
  E = M_0 + \int_{S_0}^{S} T dS = M_0 + 2\pi \int_{r_0}^{r_H} r'_H
  T(r'_H, R) dr'_H,
\end{equation}
Here, the boundary condition $E = M_0$ for $r_H = r_0$ is considered and
the thermodynamic energy of the RN black hole is 
\begin{equation}
  \label{E:RN}
  E^{\rm RN} = \tilde{M}_0 + 2\pi \int_{\tilde{r}_0}^{\tilde{r}_+}
  \tilde{r}'_+ T^{\rm RN}(\tilde{r}'_+, \tilde{Q}, \tilde{R})
  d\tilde{r}'_+, 
\end{equation}
where the local temperature of the RN black hole is written as
\begin{equation}
  \label{T:RN}
  T^{\rm RN} = \frac{T_H^{\rm RN}}{\tilde{f}(\tilde{R})}.
\end{equation}
Moreover, our definition of energy~(\ref{E})
is consistent with that of the Schwarzschild black hole in
Ref.~\cite{york} for the limit of $\theta \to 0$, and the energy is
positive definite as shown in Fig.~\ref{fig:E}.

\begin{figure}[pt]
  \includegraphics[width=0.5\textwidth]{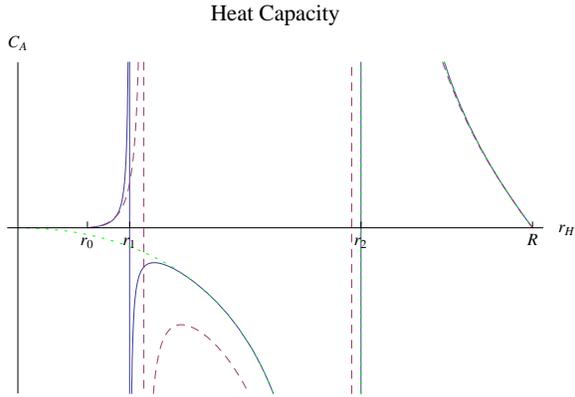}
  \caption{The solid, the dashed, and the dotted lines show the 
        relations between the local temperature and the horizon of the
        noncommutative, the scaled RN, and the Schwarzschild black
        holes, respectively.
      }
  \label{fig:C}
\end{figure}

In order to check the stability of the noncommutative black hole, we
calculate the heat capacity as
\begin{equation}
  \label{C}
  C_A = \left( \frac{\partial E}{\partial T} \right)_R, 
\end{equation}
where $A=4\pi R^2$ is the area of the boundary of the cavity.
The Fig.~\ref{fig:C} shows the behavior of the heat capacity. Since the
heat capacity is positive for $r_0 < r_H < r_1$ and $r_H >
r_2$, the small and the large black holes are stable. In the case of
$r_1 < r_H < r_2$, the black hole is unstable since the heat capacity
is negative. And the heat capacity approaches zero as $r_H$ goes to
$r_0$ or $R$. This stability can be examined by considering the free
energy.

\begin{figure}[pt]
  \includegraphics[width=0.5\textwidth]{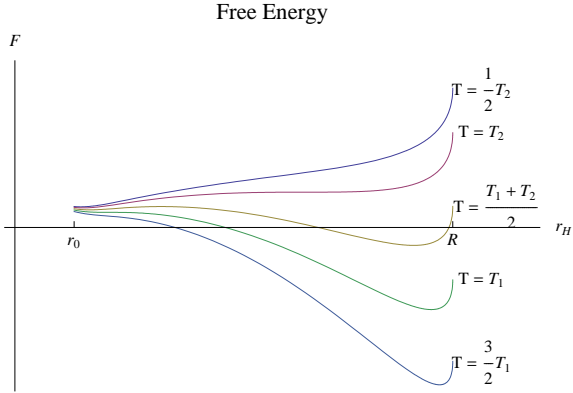}
  \caption{The free energy has one minimum for $0<T<T_2$ and $T>T_1$
    and three extrema for $T_2 < T < T_1$. The number of the extrema
    gives the number of possible black holes. The stable black
    holes appear when the free energy has the local minimum, while the
    unstable black holes appear when it has the local maximum.
    }
  \label{fig:F}
\end{figure}

The off-shell free energy of the noncommutative black hole within the
cavity is given by
\begin{equation}
  \label{F}
  F = E(r_H, R) - T S(r_H),
\end{equation}
where $E$ and $S$ are given from Eqs.~(\ref{E}) and (\ref{S}),
respectively, and $T$ is an arbitrary temperature. Fig.~\ref{fig:F}
shows the behavior of the free energy as a function of the horizon for
several temperatures. For $0<T<T_2$ there is a small stable black
hole, while there exists a large stable black hole for $T>T_1$. For
the case of $T_2 < T < T_1$, the small and the large black hole are
stable and the intermediate black hole is unstable. 
Then, the extrema of the off-shell free energy can be obtained from
\begin{equation}
  \label{F:extrema}
  \left( \frac{\partial F}{\partial r_H} \right)_{R,T} = 0,
\end{equation}
which is nothing but $T=T(r_H)$ for a given $T$, where $T(r_H)$ is the
local temperature~(\ref{T}).

\section{Discussion}
\label{sec:dis}
It is interesting to note that the pressure of the smeared object is negative
so that it can be considered ``akin'' to the cosmological constant in
de Sitter universe~\cite{nss}. In fact, the
inside of the inner horizon $r_C$ has de Sitter-like behavior, which
can be seen from the line element near the
origin~\cite{nss,anss}. However, we are interested in the fact that
the temperature of the noncommutative black hole vanishes when the horizon
radius reaches the minimal horizon~\cite{mkp}. We have
shown that the noncommutative 
black hole has an extremal behavior near the minimal mass, and all
thermodynamic quantities are similar to those of the near-extremal RN
black hole at least in the leading order. 

In connection with the relations~(\ref{rH:r+}) and (\ref{r0:Q}), the
coordinates $r$ and $\tilde r$ are connected by $r = \xi \tilde r$ and
the two redshift functions~(\ref{f}) and (\ref{metric:RN}) also have
similar form. To see this, we expand the functions in the near
extremal limit: 
\begin{eqnarray}
  \label{f:ex}
  & & f(r) \simeq \frac{\xi}{r_0^2} \left[ (r - r_0)^2 - (r_H - r_0)^2
  \right] + O(r_{(H)} - r_0)^3, \\ 
  \label{f:RN:ex}
  & & \tilde{f}(\tilde{r}) \simeq \tilde{Q}^{-2} \left[ (\tilde{r} -
    \tilde{Q})^2 - (\tilde{r}_+ - \tilde{Q})^2 \right] +
  O(\tilde{r}_{(+)} - \tilde{Q})^3. 
\end{eqnarray}
Comparing these two equations, one finds a relation
$f(r)\simeq\xi\tilde{f}(\tilde{r})$ in the leading order. Then, the
radial parts of the two metrics yield the line elements satisfying the
relation of $ds^2 \sim dr^2/f(r) \simeq \xi
d\tilde{r}^2/\tilde{f}(\tilde{r}) \sim \xi d\tilde{s}^2$, whereas the time
coordinates have the same scale, $dt=d\tilde{t}$. 

One might think that our result is a little dubious in
the sense that general
relativity breaks down and quantum gravity should be considered when the
noncommutative length is identified with the Planck
scale, although we have not fixed the noncommutative parameter as the
Plank scale in the present calculation. 
To address this issue, the total mass and the radial coordinate are redefined as
$M \to M'=M/(2\sqrt\theta)$ and $r \to r'=r/(2\sqrt\theta)$ for convenience.
Then, thermal energy~(\ref{T}) and mass~(\ref{M}) can be written as
$T=\mathcal{T}(r'_H,R')/\sqrt\theta$ and
$M=\mathcal{M}(r'_H)\sqrt\theta$, respectively, 
where $\mathcal{T}$ and $\mathcal{M}$ are independent of $\theta$.
To affect the back reaction of the geometry, the thermal energy should
be comparable to the total mass of the black hole.
So, if we assume that the locally
maximal radiated energy $E=T_1\simeq0.015/\sqrt\theta$ at
$r_H=r_1\simeq4.8\sqrt\theta$ 
in Fig.~\ref{fig:T} is assumed to be equal to the total mass
$M(r_H=r_1)\simeq2.4\sqrt\theta$, then it can be shown that
the quantum back-reaction cannot be neglected for this
non-extremal black hole since the noncommutative length should be
$\sqrt\theta\sim\ell_\mathrm{Pl}\sim10^{-34}$~cm \cite{nss}. 
This calculation has been done 
for the large cavity size $R\gtrsim100\sqrt\theta$.
Therefore, as expected, the back-reaction effect can not be
neglected when the noncommutative parameter is identified with the
Plank constant since at this scale the radiated energy is comparable
to the size of the black hole. Now, we want to investigate the above possibility 
in the near extremal limit which is a main part of the present work. 
Note that in this limit
$\mathcal{T}\sim\epsilon r_0^2/R^2$ in the leading order,
and $\mathcal{M}$ is order of 1, where $\epsilon=(r_H-r_0)/r_0\ll1$ and $R\gg r_0$.
Thus, at the Planck scale $\sqrt\theta\sim\ell_{Pl}$, 
the radiated energy is calculated as $T \sim \epsilon (r_0/R)^2 M \ll
M$. It means that the 
quantum back-reaction can be suppressed at the Plank scale of the
noncommutativity parameter as well as 
at the scale $\sqrt\theta\gg\ell_{Pl}$ since the radiated energy is
small compared to the size of the black hole.
In the end, the noncommutativity in the near extremal limit
cools down the black hole so that quantum back-reaction may be
suppressed at least in this thermodynamic analysis.
Of course, the complete answer is not yet known because of absence of
the consistent quantum gravity.

Finally, we would like to make a couple of comments why this work is
interesting. The noncommutativity appears in the D-brane in the string
theory and its intriguing spacetime structure seems to be very special. 
However, this kind of noncommutativity also appears in the model of
a very slowly moving charged particle on the constant magnetic
field~\cite{ko} and the Chern-Simon's theory~\cite{djt}. In these
regards, some noncommutative properties happen in various models. 
As we discussed so far, it appears in the near extremal RN black hole in the
thermodynamic analysis if the noncommutative parameter is
identified with the squared electric charge with some constants. 
Moreover, the metric of the near extremal noncommutative black hole has been
explicitly identified with that of the near extremal RN black hole. So,
we hope various properties of the noncommutative black hole 
along with the thermodynamic similarity can be studied in terms of
the near extremal RN black hole. Conversely speaking, the near
extremal RN black hole which has been widely studied in the black
hole physics shares some properties of the near extremal noncommutative
black hole. On the other hand, the noncommutative black hole is
nonsingular and has de Sitter-like geometry near
center~\cite{nss}. This is the key to the thermodynamic analogy between
noncommutative black holes and RN black holes in the near extremal
limit, since de Sitter-like geometry gives inner (Cauchy) horizon in
addition to the outer (event) horizon. In fact, de Sitter core can be
seen in most regular black holes~\cite{bardeen}, which have
asymptotic Schwarzschild geometry. They are also expected to be
thermodynamically similar to the RN black hole in the near extremal
limit. Furthermore, it would be interesting to compare the statistical entropy
between the noncommutative black hole and the RN black hole in the
near extremal limit.


\acknowledgments
W.\ Kim and M.\ Yoon were in part supported by the Science Research
Center Program of the Korea Science and Engineering Foundation through
the Center for Quantum Spacetime (CQUeST) of Sogang University with
grant number R11-2005-021. And W.\ Kim and E.\ J.\ Son were in part
supported by the Korea Science and Engineering Foundation 
(KOSEF) grant funded by the Korea government(MOST) 
(R01-2007-000-20062-0). 


\end{document}